\documentclass{article}

\usepackage{PRIMEarxiv}

\usepackage[utf8]{inputenc} 
\usepackage[T1]{fontenc}    
\usepackage{hyperref}       
\usepackage{url}            
\usepackage{booktabs}       
\usepackage{amsfonts}       
\usepackage{nicefrac}       
\usepackage{microtype}      
\usepackage{lipsum}
\usepackage{fancyhdr}       
\usepackage{graphicx}       
\graphicspath{{media/}}     
\usepackage{adjustbox}
\usepackage{booktabs} 
\usepackage{mwe} 
\usepackage{kotex}
\usepackage{tikz}
\usepackage{makecell}
 \usepackage{microtype}
\usetikzlibrary{positioning,arrows.meta}
\title{Chunk Knowledge Generation Model for Enhanced Information Retrieval: A Multi-task Learning Approach}

\author{
  Jisu Kim, Jinhee Park, Changhyun Jeon, Jungwoo Choi, Keonwoo Kim, Minji Hong, Sehyun Kim \\
  HANCOM / Seongnam, South Korea \\
  \texttt{\{jisu.kim, jhpark1, jeonch, jungwoo.choi, keonwoo.kim, minji.hong, sehyun.kim\}@hancom.com} \\
}

\begin{document}
\maketitle

\begin{abstract}
Traditional query expansion techniques for addressing vocabulary mismatch problems in information retrieval are context-sensitive and may lead to performance degradation. As an alternative, document expansion research has gained attention, but existing methods such as Doc2Query have limitations including excessive preprocessing costs, increased index size, and reliability issues with generated content. To mitigate these problems and seek more structured and efficient alternatives, this study proposes a method that divides documents into chunk units and generates textual data for each chunk to simultaneously improve retrieval efficiency and accuracy. The proposed "Chunk Knowledge Generation Model" adopts a T5-based multi-task learning structure that simultaneously generates titles and candidate questions from each document chunk while extracting keywords from user queries. This approach maximizes computational efficiency by generating and extracting three types of semantic information in parallel through a single encoding and two decoding processes. The generated data is utilized as additional information in the retrieval system. GPT-based evaluation on 305 query-document pairs showed that retrieval using the proposed model achieved 95.41\% accuracy at Top@10, demonstrating superior performance compared to document chunk-level retrieval. This study contributes by proposing an approach that simultaneously generates titles and candidate questions from document chunks for application in retrieval pipelines, and provides empirical evidence applicable to large-scale information retrieval systems by demonstrating improved retrieval accuracy through qualitative evaluation.
\end{abstract}


\section{Introduction}
One of the core challenges in information retrieval that has been addressed for decades is the vocabulary mismatch problem. This refers to the phenomenon where important documents fail to be retrieved because words in user queries do not match words in relevant documents. To mitigate this problem, various approaches have been proposed over decades, with query expansion techniques—which supplementarily add related words or phrases to queries—being the most widely used~\cite{carpineto2001survey, robertson2004understanding}.
Recently, document expansion techniques have gained attention as a new alternative to address vocabulary mismatch problems~\cite{mansour2024revisiting}. Document expansion improves document expressiveness by attaching additional words or sentences to documents, enabling matching in broader semantic spaces during indexing and retrieval. A representative approach, Doc2Query~\cite{nogueira2019document}, used T5-based sequence-to-sequence models to generate various queries from documents and indexed them together with documents to improve retrieval performance.
Such document expansion shows effects beyond simply adding meaningful words to documents. For example, even when specific query words appear only 1-2 times in the original text, adding 80 expansion queries can increase the frequency of those words dozens of times, significantly impacting the ranking process. Moreover, words that did not exist in the original document can be included, contributing to resolving potential vocabulary mismatch problems.
However, methods like Doc2Query have several drawbacks. First, generating dozens of queries per document requires expensive sequence generation operations, demanding substantial preprocessing time before indexing in large-scale document collections~\cite{yang2023auto}. Second, adding queries to documents increases overall index size, potentially causing retrieval delays or increased memory usage~\cite{moffat2023efficient}. Third, some generated expansion queries may contain content unrelated to the actual document, which can degrade retrieval accuracy. To address these limitations, the Doc2Query approach introduced filtering methods that remove low-scoring queries based on semantic relevance between queries and documents~\cite{gospodinov2023doc2query}.
Recognizing these limitations of existing document expansion methods, this study proposes a Chunk Knowledge Generation Model as a more structured and efficient document representation approach. The proposed Chunk Knowledge Generation Model divides documents into fixed-size chunks, then simultaneously generates keywords, titles, and candidate questions for each chunk and uses them as metadata for indexing. This is distinctive in that it can simultaneously improve retrieval precision and efficiency by utilizing meta-information with diverse semantic structures, beyond the single form of information augmentation through existing expansion queries.
Furthermore, this model adopts a T5-based multi-task learning structure that can perform multiple generations with a single encoding operation, making it advantageous in terms of computational efficiency and response speed. Thus, this study demonstrates that document expansion techniques can be refined and structured at the chunk level to simultaneously provide realistic scalability and performance improvements for large-scale information retrieval systems.
The contributions of this paper are as follows: (1) structured document expansion through chunk-level multi-task generation (titles, questions, keywords), (2) reduced preprocessing costs through single encoding-based parallel generation, (3) systematic comparison and performance/cost analysis across various vector indexing configurations.

\section{Related Work}
The emergence of pre-trained small language models (SLMs) has brought significant changes to the information retrieval field in terms of search and ranking structures~\cite{zhu2023large}. In particular, embedding-based retrieval technologies and neural network-based re-ranking techniques have moved beyond simple word matching to capture semantic similarity, leading to performance improvements in various retrieval environments~\cite{leonhardt2024efficient}. Various attempts to address vocabulary mismatch problems arising from traditional keyword matching through document or query expansion have gained attention~\cite{weller2023generative,mackie2023generative}.
Traditionally, query expansion methods that add related words or phrases to supplement short and simple user queries have been widely studied~\cite{carpineto2001survey}. However, query expansion has the disadvantage of requiring different expansion words for each query and being context-sensitive, potentially causing retrieval performance degradation when expansion is inappropriate~\cite{seo2025qa}. In contrast, document expansion techniques that supplement information to documents themselves are gradually emerging as alternatives~\cite{mansour2024revisiting}.
A representative document expansion technique is Doc2Query proposed by Nogueira and Lin. This research used sequence-to-sequence models to generate various queries based on documents, then added them to original documents to improve retrieval quality~\cite{nogueira2019document}.
Subsequently, Gospodinov et al. pointed out that among generated queries through document expansion, queries with low relevance to documents could actually hinder retrieval effectiveness. To solve this problem, they proposed document filtering techniques that evaluate relevance between queries and documents and attach only queries above a certain threshold to documents~\cite{gospodinov2023doc2query}.
Additionally, S. Jeong et al. proposed document expansion techniques using probabilistic text generation through unsupervised learning. They showed that diverse contextual information could be effectively provided to documents without separate labels to improve information retrieval efficiency~\cite{jeong2021unsupervised}.
Recently, document expansion research using large-scale generative language models has become active. Weller et al. explored the possibility of generating document expansion queries using ChatGPT but reported that it was unclear whether this was superior to existing T5-based Doc2Query methods~\cite{weller2023generative}. Lee et al. also proposed document expansion methods incorporating SLM-based pre-training and confirmed performance improvements in the Dense Passage Retrieval field~\cite{lee2023pretraining}.
Existing studies have mainly focused on document expansion through query generation and refining expansion queries through query-document relevance filtering. This study expands this direction by proposing a new document expansion method that simultaneously generates titles, related questions, and keywords at the document chunk level and utilizes them as metadata in the indexing stage to diversify retrieval signals, aiming to further enhance the efficiency and accuracy of semantic-based retrieval.

\section{Chunk Knowledge Generation Model}
\begin{figure}[ht]
\centering
\includegraphics[]{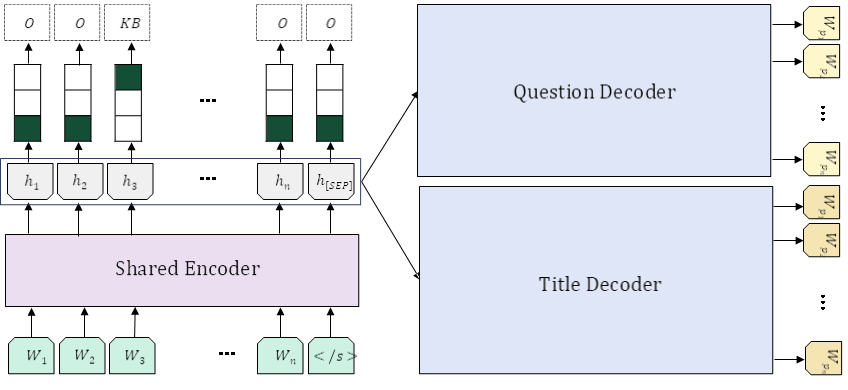}
\caption{Overall Architecture of the Proposed Chunk Knowledge Generation Model}
\label{fig:model_architecture}
\end{figure}

The Chunk Knowledge Generation Model proposed in this paper is a multi-task learning-based framework designed to maximize retrieval effectiveness at the document chunk level by simultaneously extracting and generating \textbf{keywords}, \textbf{titles}, and \textbf{candidate questions} from each chunk. The model architecture is built on the T5 family language model KETI-AIR/ke-t5-base~\cite{kim-etal-2021-model-cross,2020t5}, where the encoder is utilized for keyword extraction, while two specialized decoders independently generate titles and candidate questions from the encoder representations. After a single encoding process of the input text, the model produces task-specific outputs in parallel. The overall architecture of the proposed model is illustrated in Figure~\ref{fig:model_architecture}.

\subsection{Keyword Extraction Model}
Keyword extraction is a task for selecting meaningful central concepts from user queries. In this study, it was defined and processed as a sequence tagging problem. Specifically, a BIO tagging method similar to Named Entity Recognition (NER) was adopted, with labeling using three classes: \texttt{KB(Keyword-Begin)}, \texttt{KI(Keyword-Inside)}, and \texttt{O(Outside)}.
Input sentences are pre-separated into morpheme units using the Mecab morphological analyzer\footnote{\url{https://github.com/konlpy/konlpy}}, enabling more precise token-level prediction. For separated morphemes, it was designed as a task to classify whether each token corresponds to a keyword. The classifier is trained as a Softmax-based token classifier that takes T5 encoder output vectors as input. During training, only parts corresponding to keywords are considered as correct labels, reducing unnecessary entity learning and enabling keyword-focused concentrated learning.

\subsection{Title Generation Model}
Title generation is a task that summarizes document chunk content to describe the chunk's representative topic in one sentence. This task is defined as a text generation problem, where the decoder generates natural language sentences based on T5 encoder output vectors for input chunks, trained to generate titles limited to a maximum of one sentence length.
The decoder uses shared vectors derived from the same encoder as keyword extraction but is designed as an independent decoder structure with separate decoder parameters for each task. This allows title generation to maintain topic-centered summary generation performance without interference from other tasks.

\subsection{Candidate Question Generation Model}
Candidate question generation is a task that generates expected questions users might actually search for based on chunk content. In this study, it was configured as a task generating three natural language questions per chunk. This is also classified as a text generation task, with T5's independent decoder responsible for question generation.
The candidate question generation decoder receives shared context vectors from the encoder like title generation and is independently trained using question generation-specific parameters. This enables simultaneous output of three task results with a single encoding, providing an efficient computational structure.

\section{Experiments}

\subsection{Experimental Environment Setup}
To quantitatively verify performance improvements when titles and candidate questions generated through the Chunk Knowledge Generation Model are utilized in actual retrieval systems, comparative experiments were conducted by constructing various Vector DB configurations.

\paragraph{Embedding Model}
We used E5-large~\cite{wang2024multilingual}, a retrieval-specialized language model, was used as the embedding model. This model can effectively vectorize sentence-level semantic representations and can distinguish between sentences and questions through \textit{passage:} and \textit{query:} prompt formats. In this experiment, E5 model's special tokens \texttt{<s>} and \texttt{</s>} were utilized as delimiters to clearly distinguish between items. After combining titles and questions generated from document chunks in various ways, each item was formatted by wrapping with \texttt{<s>} … \texttt{</s>}.

\paragraph{Vector DB Environment}
We used \textbf{Qdrant}\footnote{\url{https://github.com/qdrant/qdrant}} as the vector storage supporting high-performance retrieval, configured to return document chunks most similar to queries through cosine similarity-based similarity calculation. For experiments, seven vector cases were constructed with different input configurations to compare retrieval result accuracy (Table~\ref{tab:pass_rate_comparison}).

Each case is an experimental setting to comparatively analyze the impact of Vector DB configuration methods on retrieval performance, quantitatively measuring how differences in combination order and inclusion of document chunk titles, candidate questions, and document chunks affect retrieval precision.

\paragraph{Test Environment}
Experiments were performed on NVIDIA RTX 8000 with 48GB GPU memory, and all vector processing operations including embedding inference, vector insertion, and query retrieval were repeatedly measured in the same hardware environment.

\subsection{Dataset}

\paragraph{Training and Evaluation Dataset for Chunk Knowledge Generation Model}
To train the Chunk Knowledge Generation Model that generates keywords, titles, and candidate questions at the document chunk level, diverse domain machine reading comprehension-based datasets were collected and preprocessed from four sources. The composition of each dataset is shown in Table~\ref{tab:raw_dataset}.

\begin{table}[]
\centering
\renewcommand{\arraystretch}{1.3}
\begin{tabular}{c|c|c}
\hline
\textbf{Dataset Name} & \textbf{Source} & \textbf{Sample Count} \\
\hline
Outsourcing Data & Crowdsourcing & 14,982 \\
Administrative Document MRC Data & AI-Hub & 158,546 \\
Financial/Legal Document MRC Data & AI-Hub & 123,130 \\
Book Material MRC Data & AI-Hub & 237,500 \\
\hline
\end{tabular}
\vspace{1.0em}
\caption{Original Dataset Composition}
\label{tab:raw_dataset}
\end{table}

The collected original dataset consists of 519,176 document chunks, which were directly used as units for subsequent chunk-based generation training. Considering cases where sentence length is too long to be suitable for model input, chunks exceeding 512 T5 tokens were filtered and removed. Subsequently, for filtered document chunks, the following data were generated using the \texttt{GPT-4o-mini} model: (1) one representative title summarizing the document chunk content, (2) three candidate questions derivable from the document chunk, (3) keywords derived from generated questions.

The composition scale of each item in the finally constructed dataset is shown in Table~\ref{tab:final_dataset}.

\begin{table}[]
\centering
\renewcommand{\arraystretch}{1.3}
\begin{tabular}{l|r|r|r}
\hline
\textbf{Component} & \textbf{Total} & \textbf{Validation/Test Data} & \textbf{Training Data} \\
\hline
Document Chunks & 210,522 & 20,000 & 190,522 \\
Questions & 631,566 & 60,000 & 571,566 \\
Titles & 210,522 & 20,000 & 190,522 \\
\makecell[l]{Keyword Training Queries} & 210,522 & 20,000 & 190,522 \\
\hline
\end{tabular}
\vspace{1.0em}
\caption{Final Constructed Dataset Composition and Split after Preprocessing}
\label{tab:final_dataset}
\end{table}

\paragraph{Retrieval System Evaluation Dataset}
Data for testing retrieval performance comparison were extracted from the KoRAG dataset\footnote{\url{https://huggingface.co/datasets/LDCC/korag}}. KoRAG is a question-answering dataset for Korean retrieval augmentation that searches for documents matching queries from original PDF files and generates responses based on them, composed of data with Public Nuri 1-4 type copyrights.
The original data consists of questions and answers, source PDF files for those answers, and corresponding page lists. Among these, 100 PDF files were selected for evaluation, and only questions sourcing from the selected 100 documents were extracted from the entire dataset (272 from Train data, 33 from Validation data).
To reprocess as a vector retrieval performance evaluation dataset, each PDF file was converted to text by page, with each page considered as an actual retrieval unit chunk (total 15,716 pages). To correct problems with line breaks, tables, and chart representations not being properly output during PDF text conversion, post-processing was performed using an large language model (GPT-4o-mini) to reprocess into clean chunks. The scale of the finally constructed dataset is shown in Table~\ref{tab:rag_eval_dataset}.

\begin{table}[ht]
\centering
\renewcommand{\arraystretch}{1.3}
\begin{tabular}{l|r}
\hline
\textbf{Component} & \textbf{Quantity} \\
\hline
Documents & 100 \\
Questions & 305 \\
Answer Chunks (Pages) & 338 \\
Total Chunks (Pages) & 15,716 \\
\hline
\end{tabular}
\vspace{1.0em}
\caption{Retrieval System Evaluation Dataset Composition and Quantity}
\label{tab:rag_eval_dataset}
\end{table}

\subsection{Evaluation Methods}

\paragraph{BERTScore}
BERTScore~\cite{zhang2019BERTScore} is a metric that quantitatively evaluates semantic similarity between sentences using pre-trained BERT-based language models, unlike traditional n-gram based evaluation metrics (e.g., BLEU, ROUGE). In this study, BERTScore was used to evaluate how semantically consistent generated titles, candidate questions, and keywords extracted based on questions were with references. Specifically, the google-bert/bert-base-multilingual-cased~\cite{DBLP:journals/corr/abs-1810-04805} model released by Google was used to calculate precision, recall, and F1 scores for each item. BERTScore enables more precise and reliable quality assessment for evaluating text generated or extracted based on natural language (e.g., questions, keywords) because it can consider context and meaning together, not just surface-level word matching.

\paragraph{GPT-based Evaluation}
This study conducted qualitative evaluation of overall retrieval system practical generation quality through automated evaluation prompts based on GPT-4o. This evaluation metric was applied equally to individual evaluation of title and candidate question generation as well as overall retrieval system performance. Evaluation was also conducted on whether keywords were correctly extracted from generated questions. GPT-based evaluation has been reported to show consistency and validity similar to human evaluators in evaluating natural language generation (NLG) results recently, with the advantage of more effectively evaluating contextual appropriateness or semantic delivery that are difficult to capture with quantitative metrics~\cite{gao2025llm}.
Detailed evaluation prompt specifications, exact schemas, and representative examples are provided in the appendix (Appendix~\ref{appendix:evaluation_prompt}).

\begin{itemize}
    \item \textbf{Retrieval Appropriateness Evaluation}: We evaluated whether chunks selected by the retrieval system for given questions contain key information needed to answer the corresponding questions. Evaluation criteria were based on information relevance, sufficiency, and consistency with questions. The evaluation model received questions and chunks as input, analyzed step-by-step whether the chunk contains information that can appropriately respond to the question, and finally judged as `pass' or `fail'.

    \item \textbf{Question Generation Evaluation}: We evaluated whether the questions generated for each chunk accurately reflected the information in the corresponding chunk. Evaluation criteria were whether questions accurately reflect the topic and information of the chunk and whether they were constructed without excessive inference. When multiple questions were generated for one chunk, if one or more met the criteria, it was judged as `pass'.

    \item \textbf{Title Generation Evaluation}: We evaluated whether generated titles concisely and accurately summarize key information of the corresponding chunk. The evaluators examined whether the titles faithfully represented the core content of each chunk and are expressed without information distortion as judgment criteria to determine `pass/fail'.

    \item \textbf{Question-Keyword Consistency Evaluation}: We evaluated whether keywords extracted from questions can accurately represent the meaning of the corresponding questions. As this is directly related to subsequent retrieval and filtering performance, we evaluated in a pass/fail manner depending on whether the keywords appropriately captured the key concepts of the question.
\end{itemize}

\subsection{Experimental Results}

\paragraph{BERTScore Evaluation}
 Table~\ref{tab:performance_results} shows BERTScore-based quantitative evaluation results for generated titles, candidate questions, and keywords. The proposed model achieved the highest performance with 95.0\% F1 score in title generation and also showed excellent results with 95.0\% in keyword extraction. Question generation scored relatively lower at 82.2\% F1 but maintained overall high consistency. It showed superior performance in all items compared to T5 fine-tuned models, and while Qwen3-8B and Qwen3-14B models showed some competitive results in certain items through prompt-based inference only without fine-tuning, they generally did not match the proposed model.

\begin{table}[ht]
\centering
\renewcommand{\arraystretch}{1.3}
\begin{tabular}{l|c|c|c}
\hline
\textbf{Model} & \textbf{Question} & \textbf{Title} & \textbf{Keyword} \\
\hline
\textbf{Proposed Model} & 81.1 / 83.3 / \textbf{82.2} & 95.1 / 95.0 / \textbf{95.0} & 96.6 / 93.7 / 95.0 \\
T5 fine-tuned & 81.0 / 82.8 / 81.9 & 91.9 / 91.9 / 91.9 & 92.5 / 89.7 / 91.0 \\
Qwen3-8B (prompt-only) & 79.4 / 78.3 / 78.8 & 76.1 / 85.3 / 80.4 & 89.1 / 91.3 / 90.1 \\
\textbf{Qwen3-14B (prompt-only)} & 80.5 / 80.3 / 80.4 & 78.1 / 86.1 / 81.8 & 94.9 / 95.3 / \textbf{95.1} \\
\hline
\end{tabular}
\vspace{0.5em}
\caption{Performance Comparison Based on BERTScore (\%) (Precision / Recall / F1)}
\label{tab:performance_results}
\end{table}

\paragraph{GPT-based Evaluation}
Table~\ref{tab:gpt_eval_results} shows GPT-based qualitative evaluation results for generated titles, candidate questions, and keywords. The proposed model achieved the highest accuracy of 92.0\% in title generation and also showed excellent performance of 93.6\% in keyword extraction. Question generation scored somewhat lower at 90.0\% but maintained overall balanced results. Performance improvements were achieved in all items compared to simply fine-tuned T5 models, and while Qwen3-8B and Qwen3-14B models showed competitive results in some items through prompt-based inference only due to their small language model (SLM) characteristics, the proposed model was relatively superior in overall performance and balance.

\begin{table}[h!]
\centering
\renewcommand{\arraystretch}{1.3}
\begin{tabular}{l|c|c|c}
\hline
\textbf{Model} & \textbf{Title} & \textbf{Question} & \textbf{Keyword} \\
\hline
\textbf{Proposed Model} & \textbf{92.0} & 90.0 & 93.6 \\
T5 fine-tuned & 79.6 & 80.2 & 91.8 \\
Qwen3-8B (prompt-only) & 90.4 & 95.8 & 92.4 \\
\textbf{Qwen3-14B (prompt-only)} & 91.0 & \textbf{97.6} & \textbf{94.0} \\
\hline
\end{tabular}
\vspace{0.5em}
\caption{GPT Evaluation-Based Result Comparison (Accuracy \%)}
\label{tab:gpt_eval_results}
\end{table}

\paragraph{Inference Speed and Memory Efficiency}
Results of measuring average inference time and GPU memory usage by model are shown in Table~\ref{tab:inference_speed_ratio}. All experiments were performed using NVIDIA A100 GPU (total 81,920MiB). For the proposed model, average inference time per query significantly decreases when batch size increases from 1 to 16, while GPU memory usage remains low at 6-11\% levels. This suggests the proposed model can perform fast and efficient inference even in resource-constrained environments. In contrast, SLMs like Qwen3-8B and Qwen3-14B use over 90\% of GPU memory with relatively long inference times per query.

\begin{table}[h!]
\centering
\renewcommand{\arraystretch}{1.3}
\begin{tabular}{l|c|c|c|c}
\hline
\textbf{Model} & \textbf{GPU Memory (\%)} & \textbf{Title (sec)} & \textbf{Question (sec)} & \textbf{Keyword (sec)} \\
\hline
\textbf{Proposed Model (1 batch)} & \textbf{6.87\% }& \textbf{0.1859} & \textbf{0.9544 }& \textbf{0.0094} \\
Proposed Model (16 batch) & 10.65\% & 0.0460 & 0.2633 & 0.0007 \\
T5 fine-tuned (1 batch) & 8.06\% & 0.1943 & 1.0349 & \textbf{0.0094} \\
Qwen3-8B & 90.65\% & 0.3297 & 0.9599 & 0.1787 \\
Qwen3-14B & 90.74\% & 0.6311 & 1.6304 & 0.2980 \\
\bottomrule
\end{tabular}
\vspace{1.0em}
\caption{Average inference time and GPU memory usage (\% of total) for different models.}
\label{tab:inference_speed_ratio}
\end{table}

\paragraph{Retrieval System Application Evaluation}
To compare performance of various retrieval methods, GPT-based qualitative evaluation was performed on 305 query-document pairs based on the proposed model. Retrieval accuracy evaluation was conducted using Top@1, Top@3, Top@5, Top@10 criteria, meaning whether appropriate chunks were included within the top (n) results.
According to Table~\ref{tab:pass_rate_comparison}, Cases 4-6 generally showed excellent performance, with Case 6 achieving the highest accuracy of 84.26\% at Top@1 and 95.41\% at Top@10. In contrast, Case 2, which performed full document-level retrieval, showed the lowest performance of 38.36\% at Top@1.
Even by average performance (Avg), Case 6 was most outstanding at 91.39\%, followed closely by Case 5 (91.06\%) and Case 4 (90.90\%). These results suggest that combining candidate questions and title generation data for chunks is effective for improving retrieval accuracy.

\begin{table}[ht]
\centering
\renewcommand{\arraystretch}{1.3}
\begin{tabular}{c|c|c|c|c|c}
\hline
\textbf{Case} & \textbf{Top@1} & \textbf{Top@3} & \textbf{Top@5} & \textbf{Top@10} & \textbf{Avg} \\
\hline
Case 1 & 80.00 & 89.18 & 90.82 & 92.79 & 88.19 \\
Case 2 & 38.36 & 52.13 & 55.08 & 61.31 & 51.72 \\
Case 3 & 72.46 & 88.85 & 90.49 & 92.79 & 86.14 \\
Case 4 & 83.61 & 91.48 & \textbf{94.10} & 94.43 & 90.90 \\
Case 5 & 82.62 & \textbf{92.46} & 93.77 & \textbf{95.41} & 91.06 \\
Case 6 & \textbf{84.26} & 91.80 & \textbf{94.10} & \textbf{95.41} & \textbf{91.39} \\
Case 7 & 80.66 & 87.54 & 91.48 & 92.46 & 88.03 \\
\hline
\end{tabular}
\vspace{1.0em}
\caption{Pass Rates (\%) by Various Retrieval Methods (GPT Evaluation-based)}
\label{tab:pass_rate_comparison}
\end{table}

\begin{itemize}
  \item \textbf{Case 1}: \texttt{passage: <s> document chunk </s>}
  \item \textbf{Case 2}: \texttt{passage: <s> title </s>}
  \item \textbf{Case 3}: \texttt{passage: <s> candidate questions </s>}
  \item \textbf{Case 4}: \texttt{passage: <s> title </s> candidate questions </s> document chunk </s>}
  \item \textbf{Case 5}: \texttt{passage: <s> candidate questions </s> title </s> document chunk </s>}
  \item \textbf{Case 6}: \texttt{passage: <s> candidate questions </s> document chunk </s>}
  \item \textbf{Case 7}: \texttt{passage: <s> title </s> document chunk </s>}
\end{itemize}

\section{Discussion}
While the proposed model did not achieve absolute best performance in all items compared to small language models (SLMs), it achieved balanced performance through fine-tuning. Particularly in title and keyword generation, it showed similar or better results than SLM-based prompt methods, while maintaining stable performance in question generation. This demonstrates that even with relatively small-scale models, competitive performance can be secured through appropriate fine-tuning strategies.
In terms of efficiency, the proposed model has distinct advantages over SLMs in speed and resource utilization. Experimental results showed that while Qwen3-8B and Qwen3-14B consume over 90\% of GPU memory with long inference times, the proposed model maintains low resource occupation of 6-11\% while providing fast responses. Particularly in query-keyword extraction stages, response speeds of milliseconds were recorded, greatly enhancing applicability in environments requiring real-time responsiveness. This is significant in simultaneously meeting speed and efficiency requirements that are core to RAG (Retrieval-Augmented Generation) systems.
In retrieval accuracy evaluation, the proposed model significantly improved Top@k accuracy when candidate questions and titles were combined with chunks to construct retrieval passages, with Case 6 achieving the best results of 84.26\% at Top@1 and 91.39\% average. This suggests that candidate questions and titles generated by the proposed model provide important contextual information in retrieval processes beyond simple summarization. In summary, the proposed model is not only superior in efficiency and speed compared to SLMs but also has the ability to generate auxiliary signals that enhance retrieval quality, making it a practical alternative for RAG systems.

\section{Conclusion}
This study proposed a multi-task based chunk knowledge generation model that generates titles and candidate questions from document chunks and extracts keywords from user queries. The proposed model showed high accuracy in BERTScore and GPT-based evaluations, with significant improvement in retrieval accuracy when candidate questions and titles were combined with chunks for retrieval use. The model also achieved competitive performance while maintaining significantly lower memory usage and faster inference speed compared to SLMs, proving it can be an effective alternative in RAG system environments where real-time responsiveness and resource efficiency are important. Future research plans to extend in directions that further enhance retrieval precision by combining query-based keyword extraction with metadata filtering and retrieval candidate reduction techniques.

\bibliographystyle{unsrt}  
\bibliography{references}  

\begin{thebibliography}{10}

\bibitem{carpineto2001survey}
Claudio Carpineto and Giovanni Romano.
\newblock A survey of automatic query expansion in information retrieval.
\newblock {\em ACM Computing Surveys (CSUR)}, 33(1):1--47, 2001.

\bibitem{robertson2004understanding}
Stephen Robertson.
\newblock Understanding inverse document frequency: on theoretical arguments for idf.
\newblock {\em Journal of documentation}, 60(5):503--520, 2004.

\bibitem{mansour2024revisiting}
Watheq Mansour, Shengyao Zhuang, Guido Zuccon, and Joel Mackenzie.
\newblock Revisiting document expansion and filtering for effective first-stage retrieval.
\newblock In {\em Proceedings of the 47th international ACM SIGIR conference on Research and Development in information retrieval}, pages 186--196, 2024.

\bibitem{nogueira2019document}
Rodrigo Nogueira, Wei Yang, Jimmy Lin, and Kyunghyun Cho.
\newblock Document expansion by query prediction.
\newblock {\em arXiv preprint arXiv:1904.08375}, 2019.

\bibitem{yang2023auto}
Tianchi Yang, Minghui Song, Zihan Zhang, Haizhen Huang, Weiwei Deng, Feng Sun, and Qi~Zhang.
\newblock Auto search indexer for end-to-end document retrieval.
\newblock {\em arXiv preprint arXiv:2310.12455}, 2023.

\bibitem{moffat2023efficient}
Alistair Moffat and Joel Mackenzie.
\newblock Efficient immediate-access dynamic indexing.
\newblock {\em Information Processing \& Management}, 60(3):103248, 2023.

\bibitem{gospodinov2023doc2query}
Mitko Gospodinov, Sean MacAvaney, and Craig Macdonald.
\newblock Doc2query--: when less is more.
\newblock In {\em European Conference on Information Retrieval}, pages 414--422. Springer, 2023.

\bibitem{zhu2023large}
Yutao Zhu, Huaying Yuan, Shuting Wang, Jiongnan Liu, Wenhan Liu, Chenlong Deng, Haonan Chen, Zheng Liu, Zhicheng Dou, and Ji-Rong Wen.
\newblock Large language models for information retrieval: A survey.
\newblock {\em arXiv preprint arXiv:2308.07107}, 2023.

\bibitem{leonhardt2024efficient}
Jurek Leonhardt, Henrik M{\"u}ller, Koustav Rudra, Megha Khosla, Abhijit Anand, and Avishek Anand.
\newblock Efficient neural ranking using forward indexes and lightweight encoders.
\newblock {\em ACM Transactions on Information Systems}, 42(5):1--34, 2024.

\bibitem{weller2023generative}
Orion Weller, Kyle Lo, David Wadden, Dawn Lawrie, Benjamin Van~Durme, Arman Cohan, and Luca Soldaini.
\newblock When do generative query and document expansions fail? a comprehensive study across methods, retrievers, and datasets.
\newblock {\em arXiv preprint arXiv:2309.08541}, 2023.

\bibitem{mackie2023generative}
Iain Mackie, Shubham Chatterjee, and Jeffrey Dalton.
\newblock Generative relevance feedback with large language models.
\newblock In {\em Proceedings of the 46th international ACM SIGIR conference on research and development in information retrieval}, pages 2026--2031, 2023.

\bibitem{seo2025qa}
Wonduk Seo and Seunghyun Lee.
\newblock Qa-expand: Multi-question answer generation for enhanced query expansion in information retrieval.
\newblock {\em arXiv preprint arXiv:2502.08557}, 2025.

\bibitem{jeong2021unsupervised}
Soyeong Jeong, Jinheon Baek, ChaeHun Park, and Jong~C Park.
\newblock Unsupervised document expansion for information retrieval with stochastic text generation.
\newblock {\em arXiv preprint arXiv:2105.00666}, 2021.

\bibitem{lee2023pretraining}
Jinhyuk Lee, Wonjin Yoon, Jaewoo Kang, Minbyul Hwang, and Sang-goo Lee.
\newblock Pretraining text encoders with document expansion for dense retrieval.
\newblock In {\em Proceedings of the 2023 Conference on Empirical Methods in Natural Language Processing (EMNLP)}, pages 12489--12501, 2023.

\bibitem{kim-etal-2021-model-cross}
San Kim, Jin~Yea Jang, Minyoung Jung, and Saim Shin.
\newblock A model of cross-lingual knowledge-grounded response generation for open-domain dialogue systems.
\newblock In {\em Findings of the Association for Computational Linguistics: EMNLP 2021}, pages 352--365, Punta Cana, Dominican Republic, November 2021. Association for Computational Linguistics.

\bibitem{2020t5}
Colin Raffel, Noam Shazeer, Adam Roberts, Katherine Lee, Sharan Narang, Michael Matena, Yanqi Zhou, Wei Li, and Peter~J. Liu.
\newblock Exploring the limits of transfer learning with a unified text-to-text transformer.
\newblock {\em Journal of Machine Learning Research}, 21(140):1--67, 2020.

\bibitem{wang2024multilingual}
Liang Wang, Nan Yang, Xiaolong Huang, Linjun Yang, Rangan Majumder, and Furu Wei.
\newblock Multilingual e5 text embeddings: A technical report.
\newblock {\em arXiv preprint arXiv:2402.05672}, 2024.

\bibitem{zhang2019BERTScore}
Tianyi Zhang, Varsha Kishore, Felix Wu, Kilian~Q Weinberger, and Yoav Artzi.
\newblock Bertscore: Evaluating text generation with bert.
\newblock {\em arXiv preprint arXiv:1904.09675}, 2019.

\bibitem{DBLP:journals/corr/abs-1810-04805}
Jacob Devlin, Ming{-}Wei Chang, Kenton Lee, and Kristina Toutanova.
\newblock {BERT:} pre-training of deep bidirectional transformers for language understanding.
\newblock {\em CoRR}, abs/1810.04805, 2018.

\bibitem{gao2025llm}
Mingqi Gao, Xinyu Hu, Xunjian Yin, Jie Ruan, Xiao Pu, and Xiaojun Wan.
\newblock Llm-based nlg evaluation: Current status and challenges.
\newblock {\em Computational Linguistics}, pages 1--27, 2025.

\end{thebibliography}
\appendix
\section{Evaluation Prompt Examples}
\label{appendix:evaluation_prompt}

We provide the detailed structure of the evaluation prompts used in GPT-based evaluation.

\subsection{Keyword Evaluation Prompt}
\paragraph{System prompt}
\begin{verbatim}
당신은 query를 기반으로 keyword가 잘 추출되었는지 평가하는 전문가입니다.
\end{verbatim}

\paragraph{User prompt}
\begin{verbatim}
다음은 주어진 질문을 기반으로 추출된 키워드가 적절한지 평가하기 위한 데이터입니다.
[BEGIN DATA]

[평가 기준]: 질문에 기반하여 추출된 키워드가 해당 정보를 정확하고 적절하게 반영하고 있는가
[질문]: {Question}
[키워드]: {Keywords}
[정답 키워드 (참고용)]: {Ground truth}

키워드는 반드시 질문에서 추출된 핵심 정보여야 하며, 의미가 불분명하거나 대표성이 약한 경우 false로 표기합니다.
※ 정답 질문은 단순 참고를 위한 예시일 뿐이며, 평가 근거로 사용하지 마세요.

[END DATA]

생성된 키워드가 위 기준에 부합하는지 평가하세요.  
그 후, 한 줄 아래에 'pass' 또는 'fail' 중 하나를 명확히 적으세요.  
또 한 줄 아래에 최종 판단을 다시 한 번 반복해서 적어 주세요 (따옴표 없이, 단어 그대로 작성).
\end{verbatim}

\subsection{Title Evaluation Prompt}
\paragraph{System prompt}
\begin{verbatim}
당신은 chunk를 기반으로 생성된 제목이 적절한지 평가하는 전문가입니다. 주어진 평가 기준에 따라 제목이 해당 chunk의 정보에 충실하게 생성되었는지를 신중하고 논리적으로 판단하세요.
\end{verbatim}

\paragraph{User prompt}
\begin{verbatim}
다음은 주어진 chunk를 기반으로 생성된 제목이 적절한지 평가하기 위한 데이터입니다.
[BEGIN DATA]

[평가 기준]: 문서 조각에 기반하여 생성된 제목이 해당 정보를 정확하고 적절하게 반영하고 있는가
[Chunk]: {Chunk}
[제목]: {Title}
[정답 title (참고용)]: {Ground truth}

※ 정답 title은 단순 참고를 위한 예시일 뿐이며, 평가 근거로 사용하지 마세요.
※ 평가는 반드시 [Chunk]와 [제목]만을 기준으로 수행하세요.


평가자는 아래 기준에 따라 생성된 제목이 ‘통과(pass)’인지 ‘실패(fail)’인지를 판단해야 합니다.
이 평가의 핵심은 제목의 표현 방식이나 문장 구조가 아니라, chunk에 담긴 정보를 함축한 제목인가를 평가하는 것입니다.

평가 기준은 다음과 같습니다:
- 제목이 chunk의 핵심 정보나 주제를 반영하고 있는가
- chunk의 정보만으로 질문에 대한 충분한 근거가 확보되는가
- 제목이 chunk에 포함되지 않은 내용을 추론하거나 과도하게 확장하고 있지는 않은가
- chunk의 의미를 왜곡하지 않고 정확히 반영했는가

다음과 같은 경우는 ‘실패(fail)’로 평가합니다:
- chunk의 핵심 내용과 관련 없는 제목이 생성됨
- chunk에 포함되지 않은 정보에 기반한 제목이 생성됨
- chunk의 정보로는 제목이 과도한 추론으로 구성될 수 없음
- chunk의 의미를 잘못 해석하여 왜곡된 제목이 생성됨

[END DATA]

생성된 질문이 위 기준에 부합하는지 평가하세요.
그 후, 한 줄 아래에 'pass' 또는 'fail' 중 하나를 명확히 적으세요.
\end{verbatim}

\subsection{Question Evaluation Prompt}
\paragraph{System prompt}
\begin{verbatim}
당신은 chunk를 기반으로 생성된 질문이 적절한지 평가하는 전문가입니다. 주어진 평가 기준에 따라 질문이 해당 chunk의 정보에 충실하게 생성되었는지를 신중하고 논리적으로 판단하세요.
\end{verbatim}

\paragraph{User prompt}
\begin{verbatim}
다음은 주어진 chunk를 기반으로 생성된 질문 세트가(<q1>, <q2>, <q3>) 적절한지 평가하기 위한 데이터입니다.
[BEGIN DATA]

[평가 기준]: 문서 조각에 기반하여 생성된 질문 세트가 해당 정보를 정확하고 적절하게 반영하고 있는가
[Chunk]: {Chunk}
[질문 세트]: {Questions}
[정답 질문 (참고용)]: {Ground truth}

※ 정답 질문은 단순 참고를 위한 예시일 뿐이며, 평가 근거로 사용하지 마세요.
※ 평가는 반드시 [Chunk]와 [질문 세트]만을 기준으로 수행하세요.


평가자는 아래 기준에 따라 생성된 질문이 ‘통과(pass)’인지 ‘실패(fail)’인지를 판단해야 합니다.
이 평가의 핵심은 질문의 표현 방식이나 문장 구조가 아니라, 해당 질문이 주어진 chunk의 내용으로부터 실제로 도출될 수 있는지 여부를 판단하는 것입니다.

평가 기준은 다음과 같습니다:
- 질문이 chunk의 핵심 정보나 주제를 반영하는가
- chunk의 정보만으로 질문에 대한 충분한 근거가 확보되는가
- 질문이 chunk에 포함되지 않은 내용을 추론하거나 과도하게 확장하지 않았는가
- 질문이 chunk의 의미를 왜곡하지 않았는가

다음과 같은 경우는 ‘실패(fail)’로 평가합니다:
- chunk의 내용과 관련 없는 질문이 생성됨
- chunk에 포함되지 않은 정보에 기반한 질문이 생성됨
- chunk의 정보로는 질문이 과도한 추론으로 구성될 수 없음
- chunk의 의미를 잘못 해석하여 왜곡된 질문이 생성됨

[END DATA]

생성된 질문이 위 기준에 부합하는지 평가하세요.
질문 세트 중 하나라도 chunk의 정보를 정확히 반영하면 전체 평가를 pass로 한다.
세 질문 모두 chunk와 무관하거나 부적절하다면 fail로 한다.
그 후, 한 줄 아래에 'pass' 또는 'fail' 중 하나를 명확히 적으세요.
\end{verbatim}

\subsection{Retrieval System Application Evaluation Prompt}
\paragraph{System prompt}
\begin{verbatim}
당신은 질문에 적절히 답하기 위한 정보가 포함된 문서 조각(chunk)을 평가하는 전문가입니다. 주어진 평가 기준에 따라 각 chunk가 질문에 적절한 정보를 담고 있는지 신중하고 논리적으로 판단하세요. 먼저, 평가 근거를 단계별로 한국어로 설명한 뒤, 마지막에는 'pass' 또는 'fail' 중 하나를 명확히 기재합니다. 판단 기준은 정보의 관련성과 충분성에 기반해야 하며, 응답의 표현 방식이 아니라 정보의 포함 여부에 집중해야 합니다.
\end{verbatim}

\paragraph{User prompt}
\begin{verbatim}
다음은 주어진 질문에 대해 응답을 생성하기 위해 선택된 정보(chunk)가 적절했는지를 평가하기 위한 데이터입니다.
[BEGIN DATA]

[평가 기준]: 질문에 적절히 답변하기 위해 필요한 정보를 포함하는 chunk를 올바르게 선택했는가

[질문]: {query}

[Chunk]: {chunk}

평가자는 아래 기준에 따라 선택된 chunk가 '통과(pass)'인지 '실패(fail)'인지를 판단해야 합니다. 이 평가의 핵심은 응답의 완성도나 표현이 아니라, 질문에 답하기 위해 어떤 정보가 필요했는지, 그리고 그 정보가 chunk에 포함되어 있었는지를 평가하는 것입니다.

평가 기준은 다음과 같습니다:
- 질문에 핵심적으로 필요한 정보가 chunk에 포함되어 있어야 합니다.
- chunk가 질문의 의도와 관련된 내용을 충분히 담고 있어야 합니다.
- chunk가 불필요한 정보로 구성되어 있거나, 질문과 무관한 내용을 포함해서는 안 됩니다.
- 필요한 정보가 누락되었거나, 전혀 관련 없는 정보만 포함된 경우는 실패로 간주합니다.

다음과 같은 경우에는 '실패(fail)'로 평가합니다:
- 질문에 답하기 위한 핵심 정보가 chunk에 포함되어 있지 않음
- chunk가 질문의 의도와 전혀 관련 없는 내용으로 구성됨
- chunk가 너무 단편적이거나 파편화되어 정보의 핵심을 파악할 수 없음
- chunk 없이도 답변이 가능할 정도로 비본질적인 내용만 포함됨

[END DATA]

선택된 chunk가 위 기준에 부합하는지 평가하세요. 먼저, 한국어로 평가 근거를 단계별로 설명하세요. 판단 결과는 즉시 밝히지 마세요.
그 후, 한 줄 아래에 'pass' 또는 'fail' 중 하나를 명확히 적으세요.
또 한 줄 아래에 최종 판단을 다시 한 번 반복해서 적어 주세요 (따옴표 없이, 단어 그대로 작성).
\end{verbatim}

\end{document}